\documentclass{article}
\usepackage{graphicx}
\usepackage{cite}
\begin{document}
\title{Minimal Length and Bouncing Particle Spectrum}
\author{Kourosh Nozari$^a$\thanks{knozari@umz.ac.ir} and Pouria Pedram$^b$\thanks{pouria.pedram@gmail.com}\\
{\small $^a$Department of Physics, Islamic Azad University, Sari Branch, Sari, Iran}\\
 {\small $^b$Plasma Physics Research Center, Science and Research
Branch, Islamic Azad University, Tehran, Iran}}
\date{\today}
\maketitle \baselineskip 24pt

\begin{abstract}
In this paper we study the effects of the Generalized Uncertainty
Principle (GUP) on the spectrum of a particle that is bouncing
vertically and elastically on a smooth reflecting floor in the
Earth's gravitational field (a quantum bouncer). We calculate energy
levels and corresponding wave functions of this system in terms of
the GUP parameter. We compare the outcomes of our study with the
results obtained from elementary quantum mechanics. A potential
application of the present study is discussed finally.
\end{abstract}


\textit{Pacs}: {04.60.-m}\\

\section{Introduction}
The Generalized Uncertainty Principle (GUP) is a common feature of
all promising candidates of quantum gravity. String theory, loop
quantum gravity, black hole physics, and noncommutative geometry
(based on a deeper insight to the nature of quantum spacetime at
Planck scale) all support the need for a necessary modification of
the standard Heisenberg principle. It has been shown that
measurements in quantum gravity are governed by the generalized
uncertainty principle \cite{1,2,3,4,5,6,7,8,9,10}. There are some
evidences from string theory and black holes physics (based on some
gedanken experiments), that lead authors to re-examine the usual
uncertainty principle of Heisenberg \cite{11}. These evidences have
origin on the quantum fluctuation of the background spacetime metric
and are related to the very nature of spacetime in quantum gravity
era. The introduction of this idea has drown attention and many
authors considered various problems in the framework of the
generalized uncertainty principle (see for instance
\cite{12,13,14,16,17,18,19,20,21,22,23,24,das,p1,nozari}). These
investigations have revealed some new features of the very nature of
spacetime: spacetime is not commutative at Planck scale and it has a
foam-like structure in this scale, it seems that gravity is not a
fundamental interaction of the nature (it may be induced by the
residual effects of fundamental quantum fields on the vacuum, with
the Lagrangian playing the role of an elastic stress), constants of
the nature are not really constant and the very notion of locality
in position space representation breaks down in Planck scale.
Therefore, it seems that a reformulation of quantum theory is
required in order to incorporate gravitational effects in Planck
scale phenomena. These issues have been the topics of a wide range
of researches in recent years.

In this paper, we consider the problem of a particle of mass $m$
that is bouncing vertically and elastically on a smooth reflecting
floor in Earth's gravitational field. We solve this problem in the
presence of a minimal length within the GUP framework. First, we
give an overview to the GUP formalism and in this way we obtain a
generalized Schr\"odinger equation. After studying this equation in
the momentum space, we find the modified eigenstates and energy
spectrum of this system. We compare the outcomes of our study with
the results obtained from elementary quantum mechanics. A potential
application of the present study in the spontaneous decay of an
excited state for Ultra Cold Neutrons bouncing above a perfect
mirror in the Earth's gravitational field is discussed finally. We
note that modification to the decay rate due to existence of a
minimal length studied here, becomes important at or above the
Planck energy. Although this modification is too small to be
measurable at present, we speculate on the possibility of extracting
measurable predictions in the future.

\section{A Generalized Uncertainty Principle}
Quantum mechanics with modification of the usual canonical
commutation relations has been investigated intensively in the last
few years (see \cite{26} and references therein). Such works which
are motivated by several independent streamlines of investigations
in string theory and quantum gravity, suggest the existence of a
finite lower bound to the possible resolution $\Delta X$ of
spacetime points. The following deformed commutation relation has
attracted much attention in recent years \cite{26}
\begin{equation}\label{com}
[X, P]=i\hbar(1+\beta P^2),
\end{equation}
and it was shown that it implies the existence of a minimal
resolution length $\Delta X=\sqrt{\langle X^2 \rangle -\langle X
\rangle^2}\ge\hbar\sqrt\beta$ \cite{26}. This means that there is no
possibility to measure coordinate $X$ with accuracy smaller than
$\hbar\sqrt\beta$. Since in the context of the string theory the
minimum observable distance is the string length, we conclude that
$\sqrt{\beta}$ is proportional to this length. If we set $\beta=0$,
the usual Heisenberg algebra is recovered. The use of the deformed
commutation relation (\ref{com}) brings new difficulties in solving
the quantum problems. A part of difficulties is related to the break
down of the notion of locality and position space representation in
this framework \cite{26}. The above commutation relation results in
the following uncertainty relation
\begin{eqnarray}\label{gup}
 \Delta X \Delta P \geq \frac{\hbar}{2}
\left( 1 +\beta (\Delta P)^2 +\gamma \right),
\end{eqnarray}
where $\beta$ and $\gamma$ are positive quantities which depend on
the expectation value of the position and the momentum operators. In
fact, we have $\beta=\beta_0/(M_{Pl} c)^2$ where $M_{Pl}$ is the
Planck mass and $\beta_0$ is of the order of unity. We expect that
these quantities are only relevant in the domain of the Planck
energy $M_{Pl} c^2\sim 10^{19}GeV$. Therefore, in the low energy
regime, the parameters $\beta$ and $\gamma$ are irrelevant and we
recover the well-known Heisenberg uncertainty principle. These
parameters, in principle, can be obtained from the underlying
quantum gravity theory such as string theory.

Note that $X$ and $P$ are symmetric operators on the dense domain
$S_{\infty}$ with respect to the following scalar product \cite{26}
\begin{eqnarray}
\langle\psi|\phi\rangle=\int_{-\infty}^{+\infty}\frac{dP}{1+\beta
P^2}\psi^{*}(P)\phi(P).
\end{eqnarray}
Moreover, the comparison between Eqs.~(\ref{com}) and (\ref{gup})
shows that $\gamma=\beta\langle P\rangle^2$. Now, let us define
\begin{eqnarray}\label{x0p0}
\left\{
\begin{array}{ll}
X = x,\\\\ P = p \left( 1 + \frac{1}{3}\beta\, p^2
\right),
\end{array}
\right.
\end{eqnarray}
where $x$ and $p$ obey the canonical commutation relations
$[x,p]=i\hbar$. One can check that using Eq. (\ref{x0p0}), Eq.
(\ref{com}) is satisfied to ${\cal{O}}(\beta)$. Also, from the above
equation we can interpret $p$ as the momentum operator at low
energies ($p=-i\hbar \partial/\partial{x}$) and $P$ as the momentum
operator at high energies. Now, consider the following form of the
Hamiltonian:
\begin{eqnarray}
H=\frac{P^2}{2m} + V(x),
\end{eqnarray}
which using Eq.~(\ref{x0p0}) can be written as
\begin{eqnarray}
H=H_0+\beta H_1+{\cal{O}}(\beta^2),
\end{eqnarray}
where $H_0=\frac{\displaystyle p^2}{\displaystyle2m} + V(x)$ and
$H_1=\frac{\displaystyle p^4}{\displaystyle3m}$.

In the quantum domain, this Hamiltonian results in the following
generalized Schr\"odinger equation in the quasi-position
representation
\begin{eqnarray}\label{H}
-\frac{\hbar^2}{2m}\frac{\partial^2\psi(x)}{\partial
x^2}+\beta\frac{\hbar^{4}}{3m}\frac{\partial^{4}\psi(x)}{\partial
x^{4}} +V(x)\psi(x)=E\psi(x),
\end{eqnarray}
where the second term is due to the generalized commutation relation
(\ref{com}). This equation is a $4$th-order differential equation
which in principle admits $4$ independent solutions. Therefore,
solving this equation in $x$ space and separating the physical
solutions is not an easy task. In the next section, for the case of
a Bouncing Particle, we find the energy spectrum and the
corresponding eigenstates up to the first order of the GUP
parameter.

\section{Spectrum of a Quantum Bouncer in the GUP Scenario}
\begin{figure}
\centering
\includegraphics[width=8cm]{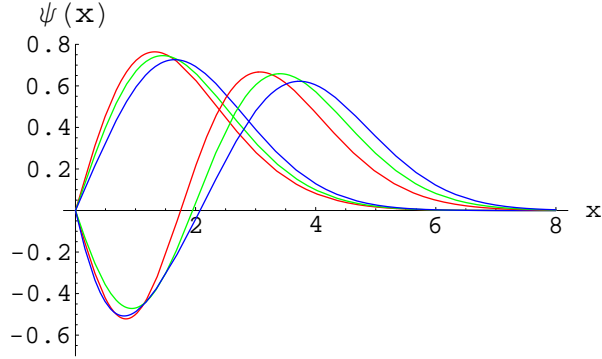} \caption{The normalized ground state and first
excited state eigenfunctions of a bouncing particle in the framework
of the generalized commutation relation (\ref{com}) for $\beta=0$
(red), $\beta=0.1$ (green), and $\beta=0.2$ (blue).} \label{fig1}
\end{figure}

Consider a particle of mass $m$ which is bouncing vertically and
elastically on a reflecting hard floor so that
\begin{eqnarray}
V(X)=\left\{
\begin{array}{cc}
mgX&\mbox{if}\quad X>0,\\\\ \infty\,
&\mbox{if}\quad X\leq 0,
\end{array}
\right.
\end{eqnarray}
where $g$ is the acceleration in the Earth's gravitational field.
The Hamiltonian of the system is
\begin{equation}
H=\frac{P^2}{2m}+mgX,
\end{equation}
which results in the following generalized Schr\"odinger equation
\begin{eqnarray}
-\frac{\hbar^2}{2m}\frac{\partial^2\psi(x)}{\partial
x^2}+\beta\frac{\hbar^{4}}{3m}\frac{\partial^{4}\psi(x)}{\partial
x^{4}} +mgx\psi(x)=E\psi(x).
\end{eqnarray}
This equation, for $\beta=0$ is exactly solvable and the solutions
can be written in the form of the Airy functions. Moreover, the
energy eigenvalues are related to the zeros of the Airy function.
However, for $\beta\ne0$, the situation is quite different. Because,
we need to solve a forth order differential equation and eliminate
the unphysical solutions. On the other hand, because of the linear
form of the potential, this equation can be cast into a first order
differential equation in the momentum space. Since the later form is
much easier to handle, we define a new variable $z=x-\frac{E}{mg}$ and
rewrite above equation in the momentum space, namely
\begin{equation}
\frac{p^2}{2m}\phi(p)+\beta\frac{p^4}{3m}\phi(p)+i\hbar mg\phi'(p)=0,
\end{equation}
where $\phi(p)$ is the inverse Fourier transform of $\psi(z)$ and
the prime denotes the derivative with respect to $p$. It is
straight forward to check that this equation admits the following
solution
\begin{equation}
\phi(p)=\phi_0\exp\left[\frac{i}{6m^2g\hbar}\left(p^3+\frac{2\beta}{5}
p^5\right)\right].
\end{equation}
Since $\beta$ is a small quantity, we can expand the above solution
up to the first order of $\beta$ as
\begin{equation}
\phi(p)\simeq\phi_0\exp\left({\frac{ip^3}{6m^2g\hbar}}\right)\left(1+\frac{i\beta
p^5}{15m^2g\hbar}+{\cal{O}}(\beta^2)\right).
\end{equation}
Now, using the Fourier transform, we can find the solution in the
position space up to a normalization factor
\begin{eqnarray}
\psi(x)&=&\mbox{Ai}\left[\alpha\left(x-\frac{E}{mg}\right)\right]+\frac{4}{15}\beta
m^2g\left(x-\frac{E}{mg}\right)\nonumber\\
&\times&\left\{4\mbox{Ai}\left[\alpha\left(x-\frac{E}{mg}\right)\right]+
\left(x-\frac{E}{mg}\right)\right.\nonumber\\
&\times&\left.\mbox{Ai}'\left[\alpha\left(x-\frac{E}{mg}\right)\right]\right\},
\end{eqnarray}
where $\alpha=\left(\frac{2m^2g}{\hbar^2}\right)^{1/3}$ and the
prime denotes derivative with respect to $x$. Finally, since the
potential is infinite for $x\le0$, we demand that the wave function
should vanish at $x=0$. This condition results in the quantization
of the particle's energy, namely
\begin{eqnarray}\label{zero}
\mbox{Ai}\left(-\frac{\alpha E_n}{mg}\right)-\frac{4}{15}\beta
mE_n\left[4\mbox{Ai}\left(-\frac{\alpha E_n}{mg}\right)\right.\nonumber\\
\left.-\frac{E_n}{mg}\mbox{Ai}'\left(-\frac{\alpha
E_n}{mg}\right)\right]=0.
\end{eqnarray}
To proceed further and for the sake of simplicity, let us work in the
units of $g=2\hbar=4m=2$. In this set of units, the energy
eigenvalues are the minus of the roots of the following algebraic
equation
\begin{equation}\label{root}
\mbox{Ai}\left(x\right)+\frac{2}{15}\beta
x\left[4\mbox{Ai}\left(x\right)+ x\mbox{Ai}'\left(x\right)\right]=0.
\end{equation}
So, the energy eigenvalues will be quantized and result in the following eigenfunctions
\begin{eqnarray}
\psi_n(x)&=&\mbox{Ai}\left(x-E_n\right)+\frac{2}{15}\beta
(x-E_n)\left[4\mbox{Ai}\left(x-E_n\right)\right.\nonumber\\ &+& \left.(x-E_n)\mbox{Ai}'\left(x-E_n\right)\right],
\end{eqnarray}
where $E_n$\,s should satisfy Eq.~(\ref{root}). Figure \ref{fig1}
shows the resulting normalized ground state and first excited state
eigenfunctions for $\beta=0\,,0.1,\,0.2$. Moreover, the calculated
values of the energy eigenvalues for the first ten states are also
shown in Table \ref{tab1}. These results show that the presence of
$\beta$ increases the energy levels in agreement with the functional
form of $H_1$. In other words, existence of a minimal length results
in a positive shift in the energy levels of quantum bouncer.

\section{A potential Application: Transition Rate of a Quantum Bouncer}
As a potential application of our analysis, we note that
quantization of the energy of \emph{Ultra Cold Neutrons} bouncing
above a mirror in the Earth's gravitational field has been
demonstrated in an experiment few years ago \cite{29}. This effect
demonstrates quantum behavior of the gravitational field if we
consider the spontaneous decay of an excited state in this
experiment as a manifestation of the Planck-scale effect \cite{30}.
Since the spectrum of a quantum bouncer changes in the presence of
the minimal length, we expect the rate of this decay will change as
a trace of quantum gravitational effects via existence of a minimal
length scale. In fact, as we have shown, the energy levels of a
quantum bouncer in the GUP framework attain a positive shift as
given by equation (\ref{zero}). The energy levels of a quantum bouncer in
ordinary quantum mechanics are given by zeros of Airy function as
$\lambda_{n}\approx \left(\frac{3\pi}{8}(4n-1)\right)^{2/3}$. In the presence
of a minimal length, the locations of zeros are given by
$\lambda_{n}^{(\mbox{\footnotesize{GUP}})}=\lambda_{n}+\Delta\lambda_{n}$. Within a
semi-classical analysis, we evaluate the rate for a bouncer to make
a transition $k\longrightarrow n$. The quantum quadrupole moment for
the transition  $k\longrightarrow n$ for a quantum bouncer of mass
$m$ is given by $Q_{kn}= m\langle k|X^{2}|n\rangle$ \cite{30}. The quantum
mechanical transition rate is (in the quadrupole approximation)
\begin{equation}
\Gamma_{k\longrightarrow
n}=\frac{4}{15}\frac{\omega^{5}_{kn}}{M^{2}_{Pl}c^{4}}Q^{2}_{kn},
\end{equation}
where $\omega_{kn}=(E_{k}-E_{n})/\hbar$. Now, the quadrupole matrix
element for quantum bouncer in the presence of minimal length can be
calculated using the generalized Airy function zeros given by
equation (\ref{zero}). The transition probability in our framework
is therefore
\begin{eqnarray}
\Gamma^{(\mbox{\footnotesize{GUP}})}_{k\longrightarrow
n}&=&\frac{512}{5}\frac{\left(\lambda^{(\mbox{\footnotesize{GUP}})}_{k}-
\lambda^{(\mbox{\footnotesize{GUP}})}_{n}\right)^5}{\left(\lambda_{k}-
\lambda_{n}\right)^{8}}\left(\frac{m}{M_{Pl}}\right)^{2}\frac{E^{5}_{0}c}{\alpha^{4}(\hbar
c)^{5}},\nonumber \\
&=&\left(\Gamma_{k\longrightarrow n}\right)\left(1+\frac{5\Delta
\lambda_{kn}}{\lambda_{k}- \lambda_{n}}\right)
\end{eqnarray}
where $E_{0}=mg/\alpha$, $\Delta \lambda_{kn}=\Delta\lambda_{k}-\Delta\lambda_{n}$, and $\Gamma_{k\longrightarrow
n}=\frac{512}{5}\left(\frac{m}{M_{Pl}}\right)^{2}\frac{E^{5}_{0}c}{\alpha^{4}(\hbar
c)^{5}}$. So, there will be an essentially measurable difference in the transition
rate of a quantum bouncer due to the presence of the extra $\Delta
\lambda_{kn}$ in comparison with the case that we consider just the
ordinary Heisenberg uncertainty relation.

\begin{table}
\caption{The first ten quantized energies of a bouncing particle in
GUP formalism.}\label{tab1}
\begin{center}
\begin{tabular}{cccc}
n& $\beta=0$ & $\beta=0.1$ & $\beta=0.2$ \\
\hline    0& 2.338&  2.428 & 2.570\\
 1& 4.088&  4.380 & 4.644  \\
  2& 5.521& 5.947  & 6.107  \\
  3& 6.787 &  7.257& 7.352  \\
 4& 7.944&  8.420&  8.483  \\
  5& 9.023&  9.493 & 9.536   \\
  6&10.040&  10.499&  10.532 \\
 7& 11.008&  11.456&  11.481    \\
   8&11.936& 12.371& 12.391    \\
 9&12.829& 13.253&  13.269
\end{tabular}
\end{center}
\end{table}

\section{Conclusions}
In this paper, we considered the problem of a bouncing particle in a
constant gravitational field in the framework of the generalized
uncertainty principle. We found the modified Hamiltonian and the
generalized Schr\"odinger equation as a forth order differential
equation. We solved this equation in the momentum space and obtained
the corresponding energy eigenvalues and eigenstates up to the first
order of the GUP parameter. As we have expected, we found a positive
shift in the energy spectrum due to the generalized commutation
relation. A potential application of this analysis for transition
rate of an Ultra Cold Neutron bouncing above a mirror in the Earth's
gravitational field has been explained. We emphasize that
modification to the transition rate of a quantum bouncer due to
existence of a minimal length becomes important at or above the
Planck energy. Although this modification is too small to be
measurable at present, we speculate on the possibility of extracting
measurable predictions in the future. At that case, this may provide
a direct test of underlying quantum gravity scenario.


\begin{thebibliography}{99}
\bibitem{1}
D. J. Gross and P. F. Mende, Nucl. Phys. B  {\bf 303}, 407 (1988).

\bibitem{2}
D. Amati, M. Ciafaloni and G. Veneziano, Phys. Lett. B  {\bf 216},
41 (1989).

\bibitem{3}
T. Yoneya, Mod. Phys. Lett. A  {\bf 4}, 1578 (1989).

\bibitem{4}
K. Konishi, G. Paffuti and P. Provero, Phys. Lett. B {\bf 234}, 276
(1990).

\bibitem{5}
C. Rovelli, Living Rev. Rel. {\bf 1}, 1 (1998).

\bibitem{6}
M. R. Douglas and N. A. Nekrasov, Rev. Mod. Phys.  {\bf 73}, 977
(2001).

\bibitem{7}
T. Thiemann, Lect. Notes Phys.  {\bf 631}, 41 (2003).

\bibitem{8}
A. Perez, Class. Quant. Grav. {\bf 20}, R43 (2003).

\bibitem{9}
A. Ashtekar and J. Lewandowski, Class. Quant. Grav. {\bf 21}, R53
(2004).

\bibitem{10}
F. Girelli, E. R. Livine and D. Oriti, Nucl. Phys. B  {\bf 708}, 411
(2005).

\bibitem{11}
G. Venziano,  Europhys. Lett. \textbf{2}, 199 (1986); D. Amati, M.
Ciafaloni, and G. Veneziano, Phys. Lett. B \textbf{197}, 81 (1987);
Int. J. Mod. Phys. A \textbf{3}, 1615 (1988); Phys. Lett. B
\textbf{216}, 41 (1989); Nucl. Phys. B \textbf{347}, 530 (1990); D.
J. Gross and P. F. Mende, Phys. Lett. B \textbf{197}, 129 (1987);
Nucl. Phys. B \textbf{303}, 407 (1988); K. Konishi, G. Paffuti, and
P. Provero, Phys. Lett. B \textbf{234}, 276 (1990); R. Guida, K.
Konishi, and P. Provero, Mod. Phys. Lett. A \textbf{6}, 1487 (1991);
L. J. Garay, Int. J. Mod. Phys. A \textbf{10}, 145 (1995).

\bibitem{12}
M. Maggiroe, Phys. Lett. B \textbf{304}, 65 (1993),
arXiv:hep-th/9301067 (1993).

\bibitem{13}
C. Castro, Found. Phys. Lett. \textbf{10} 273 (1997),
arXiv:hep-th/9512044.

\bibitem{14}
A. Camacho, Gen. Rel. Grav. \textbf{34}, 1839 (2002),
arXiv:gr-qc/0206006.


\bibitem{16}
S. Capozziello, G. Lambiase, G. Scarpetta, Int. J. Theor. Phys.
\textbf{39}, 15 (2000), arXiv:gr-qc/9910017.

\bibitem{17}
M. Maggiore, Phys. Rev. D \textbf{49}, 5182 (1994),
arXiv:hep-th/9305163.

\bibitem{18}
M. Maggiore, Phys. Lett. B \textbf{319}, 83 (1993),
arXiv:hep-th/9309034.

\bibitem{19}
R. J. Adler, P. Chen, D. I. Santiago, Gen.  Rel. Grav. \textbf{33},
2101 (2001).

\bibitem{20}
S. Kalyana Rama, Phys.Lett. B \textbf{519}, 103 (2001),
arXiv:hep-th/0107255.

\bibitem{21}
A. Camacho, Gen. Rel. Grav. \textbf{35}, 1153 (2003),
arXiv:gr-qc/0303061.

\bibitem{22}
P. Chen, R. J. Adler, Nucl. Phys. Proc. Suppl. \textbf{124}, 103
(2003), gr-qc/0205106.

\bibitem{23}
F. Scardigli, R. Casadio, Class.Quant.Grav. \textbf{20}, 3915
(2003), hep-th/0307174.

\bibitem{24}
A. Camacho, Rel. Grav. Cosmol. \textbf{1} 89 (2004),
arXiv:gr-qc/0302096.


\bibitem{das}S. Das and E. C. Vagenas, Can. J. Phys. \textbf{87},
233 (2009); Phys. Rev. Lett. \textbf{101}, 221301 (2008); A. F. Ali,
S. Das and E. C. Vagenas, Phys. Lett. B \textbf{678}, 497 (2009).


\bibitem{p1}P. Pedram, Europhys. Lett. \textbf{89}, 50008 (2010); Int. J. Mod. Phys.
D \textbf{19}, 2003 (2010).


\bibitem{nozari} K. Nozari and T. Azizi, Gen. Rel. Grav.
\textbf{38}, 735 (2006), K. Nozari, Phys. Lett. B \textbf{629}, 41
(2005).

\bibitem{26}
 A. Kempf, G. Mangano, and R. B. Mann, Phys. Rev. D {\bf 52}, 1108 (1995).

\bibitem{29}V. V. Nesvizhevsky \emph{et al.}, Nature \textbf{415}, 297
(2002); Phys. Rev. D \textbf{67}, 102002 (2003); Eur. Phys. J. C
\textbf{40}, 479 (2005).

\bibitem{30}G. Pignol, K. V. Ptotasov, and V. V. Nesvizhevsky, Class. Quant.
Grav. \textbf{24}, 2439 (2007).

\end{thebibliography}
\end{document}